\documentclass[twoside,english]{article}
\usepackage[T1]{fontenc}
\usepackage[latin9]{inputenc}
\usepackage{geometry}
\usepackage{fancyhdr}
\usepackage{textcomp}
\usepackage{amstext}
\usepackage{amssymb}
\usepackage{graphicx}
\PassOptionsToPackage{normalem}{ulem}
\usepackage{ulem}

\makeatletter

\DeclareRobustCommand{\greektext}{%
  \fontencoding{LGR}\selectfont\def\encodingdefault{LGR}}
\DeclareRobustCommand{\textgreek}[1]{\leavevmode{\greektext #1}}
\DeclareFontEncoding{LGR}{}{}
\DeclareTextSymbol{\~}{LGR}{126}
\newcommand{\lyxmathsym}[1]{\ifmmode\begingroup\def\b@ld{bold}
  \text{\ifx\math@version\b@ld\bfseries\fi#1}\endgroup\else#1\fi}

\makeatother

\usepackage{babel}
\begin{document}
\fancyhead{}
\fancyhead[RO]{Georgakaki D. et al} 
\fancyhead[RE]{Spectral Analysis and Allan Variance Calculation in the case of Phase Noise}

\title{Spectral Analysis and Allan Variance Calculation in the case of Phase
Noise}

\author{\uline{Georgakaki }D.$ $$^{1}${*}, Mitsas C.$ $$^{2}${*}{*},
Polatoglou H.M.$ $$ $$^{1}${*}{*}{*}}

\maketitle
\noindent \begin{center}
$^{1}$Physics Department, Solid State Physics Section, Aristotle
University of Thessaloniki., 54124, Thessaloniki, Greece 
\par\end{center}

\noindent \begin{center}
$^{2}$Mechanical Measurements Dept., Hellenic Institute of Metrology,
57022, Thessaloniki, Greece
\par\end{center}

\begin{center}
{*}georgakd@auth.gr, {*}{*}chris.mitsas@eim.gr, {*}{*}{*}hariton@auth.gr
\par\end{center}
\begin{abstract}
In this work, time series analysis techniques are used to analyze
sequential, equispaced mass measurements of a Si density artifact,
collected from an electromechanical transducer. Specifically, techniques
such as Power Spectral Density, Bretthorst periodogram, Allan variance
and Modified Allan variance can provide much insight regarding the
stochastic correlations that are induced on the outcome of an experiment
by the measurement system and establish criteria for the limited use
of the classical variance in metrology. These techniques are used
in conjunction with power law models of stochastic noise in order
to characterize time or frequency regimes by pointing out the different
types of frequency modulated (FM) or phase modulated (PM) noise. In
the case of phase noise, only Modified Allan variance can tell between
white PM and flicker PM noise. Oscillations in the system can be detected
accurately with the Bretthorst periodogram. Through the detection
of colored noise, which is expected to appear in almost all electronic
devices, a lower threshold of measurement uncertainty is obtained
and the white noise model of statistical independence can no longer
provide accurate results for the examined data set.

\end{abstract}

\section{Introduction}

Finding correlations in time series of measurements and defining the
type of stochastic noise apparent in the measured signal remain very
crucial issues for modern metrology. According to Zhang, when measurements
are autocorrelated, the use of classical variance for the estimation
of type A uncertainty of a data set, leads to wrong results {[}1{]}.
Moreover, power law models of the type 1/f$^{a}$$ $ {[}2{]} are
necessary to describe the type of stochastic noise that characterizes
the system, which produces the time series.

In most cases, due to colored noise generation, the Frequency Modulated
White Noise model (White-FM) fails to describe the system. An additional
problem is Phase Modulated Noise (PM), apparent in all electrical
oscillators {[}3{]} and more complex devices, as thoroughly reviewed
and analyzed by Rubiola {[}4{]}. PM Noise is attributed to the phase
fluctuations (or frequency fluctuations) of a signal produced by a
real, non-ideal oscillator. In this oscillator the frequency of oscillation
is modulated by electronic noise, inevitably leading to random frequency
fluctuations in time and thus affecting the performance of the system
that uses the oscillator. A list of reasons for the PM Noise generation
is given by Howe {[}5{]} but still its origin remains a controversy
issue in the areas of metrology.

Generally there have been utilized several methods from both time
and frequency domain, in order to best analyze the response of metrological
instruments, electromechanical transducers in our case study. In this
paper we will focus on Allan and Modified Allan variance (time domain)
as well as Schuster and Bretthorst periodograms (frequency domain). 

Allan variance, initially introduced by Allan in time and frequency
metrology {[}6{]}, is an alternative approach for the variance calculation
of autocorrelated measurements and has been thoroughly used by Witt
in the area of electrical measurements {[}7{]}. Because of the fact
that this method cannot tell the difference between White PM Noise
and Flicker PM Noise, the Modified Allan variance has been used instead
{[}8{]}.

In the frequency domain, Fast Fourier Transform and Schuster periodograms
remain the most widely used tools for evaluating the different types
of noise in the signal\textquoteright{} s Power Spectrum. Schuster
periodograms can detect oscillations in the signal and interpret them
as individual peaks. Due to the limitations and in certain occasions
misleading results of the Discrete Fourier Transform, an alternative
method is proposed, based on Bayesian analysis: the Bretthorst periodogram
{[}9{]}. This method also identifies oscillations in a signal and
can distinguish between two very close frequency peaks that FFT treats
as one.

The data on which the above methods will be tested, was obtained from
characterization experiments of a newly commissioned 1kg/ 10mg resolution
mass comparator at the Hellenic Institute of Metrology density laboratory.
The principle function of this comparator will be the measurement
of mass in air of Si and ceramic density artifacts of mass ca. 1kg
in the form of spheres through comparative weighing. It should be
noted that these methods can be applied to any electromechanical transducer
system equally well with similar results.

\section{General Considerations}

Let\textquoteright{} s consider an input signal of the following simple
sinusoidal form

\begin{center}
\begin{equation}
V_{in}(t)=V_{c}sin(2\pi f_{c}t)
\end{equation}

\par\end{center}

The output of a non-ideal oscillator will be of the form

\begin{center}
\begin{equation}
V_{out}(t)=(V_{c}+\varepsilon(t))sin(2\pi f_{c}+\phi(t))
\end{equation}

\par\end{center}

\noindent where the time variations in amplitude are included into
\textgreek{e}(t) and the frequency deviation from the fundamental
one is described by \textgreek{f}(t), called the phase fluctuations
of the output signal. Ignoring amplitude variations and changing to
frequency regime, phase fluctuations with a spectral density S$_{\phi}$$ $(f)
and frequency fluctuations with a spectral density S$_{y}$(f), are
described with equations (3) and (4), with h$_{i}$ representing the
intensity coefficients of each power law noise contribution and $\alpha,$
$\beta$ are the spectral exponents:

\begin{center}
\begin{equation}
S_{y}(f)={\displaystyle \sum_{a=-2}^{+2}h_{a}f^{a}}
\end{equation}

\par\end{center}

\begin{center}
\begin{equation}
S_{\phi}(f)={\displaystyle \sum_{\beta=-4}^{0}h_{\beta}f^{\beta},\;}\beta=a-2
\end{equation}

\par\end{center}

\section{Computational Methods}

\subsection{Fourier Spectral Analysis}

The Discrete Fourier Transform is one of the most powerful tools in
signal analysis and thoroughly used in many physical systems. The
periodogram was introduced by Schuster, as a method of detecting periodicities
and estimating their frequencies and is essentially the squared magnitude
of the discrete fourier transform of the data. In 1965, Cooley and
Tukey introduced the Fast Fourier Transform, a method that led to
more efficient computer calculations and is considered nowadays an
optimal frequency and power spectral estimator.

By plotting S(f) - f in a log-log diagram, the different noise levels
in the power spectrum can be estimated from the adjustment of equation
(3) to the log-log plot. The calculation of the spectral exponent
\textgreek{a}, leads to the identification of white noise (\textgreek{a}=0),
flicker noise (\textgreek{a}=-1), random walk noise (\textgreek{a}=-2)
or any intermediate case -2<\textgreek{a}<2 of colored noise {[}7{]}.

Unfortunately, the Discrete Fourier Transform will provide accurate
frequency estimates only when the following conditions are met {[}9,10{]}:
1) the length of the data series N is sufficiently large, 2) the data
series is stationary, 3) there is no evidence of low frequency existence,
4) the data contain one stationary frequency, 5) the noise is white.
When one or more of the above conditions is violated (the last one
in our work) an alternative method is proposed, based on Bayesian
Analysis, called the Bretthorst Periodogram.

\subsection{Bretthorst Periodogram}

Bayesian Probability Theory defines a probability of occurrence as
a reasonable degree of belief, given some prior information of the
data. In our example the probability distribution of a frequency of
oscillation \textgreek{w} is computed conditional on the data D and
the prior information I, abbreviated as the posterior probability
P(\textgreek{w}|D,I). To calculate this one must find the direct probability
(or likelihood function) P(D|\textgreek{w},\textgreek{I}), the prior
probability P(\textgreek{w}|\textgreek{I}), the probability P(D|I)
(in parameter estimation problems it is a simple normalization constant)
and eliminate the nuisance parameters. 

\begin{equation}
P(\omega|D,I)=\frac{P(D|\omega,I)\centerdot P(\omega|I)}{P(D|I)}
\end{equation}

Based simply on Bayes product and sum rules {[}11{]} one can obtain
the posterior probability of the frequency with or without prior knowledge
of the noise variance $\sigma^{2}$ in the signal, concluding to equations
(6a) and (6b) correspondingly. Specifically (6b) is the well-known
t-student distribution. An analytical proof of these relations can
be found in {[}9, 10{]}. C($\omega$) is the Schuster periodogram
and $\bar{d^{2}}$ is the observed mean-square data value.

$ $
\begin{equation}
\begin{array}{cccc}
P(\omega|D,\sigma,I)\varpropto exp(C(\omega)/\sigma^{2}) & (a) & P(\omega|D,I)\propto[1-\frac{2C(\omega)}{N\overline{d^{2}}}] & (b)\end{array}
\end{equation}

Basically, the above equations give evidence that a single stationary
frequency is present in the data and because of the processing in
equations (6a-6b), all details of the periodogram are suppressed except
the frequency peak, observed as a single spike in the Bretthorst periodogram.

\subsection{Allan Variance}

The Allan variance or Two-sample variance was first introduced by
David W. Allan for the evaluation of the stability of time and frequency
standards {[}6{]}. As well established, the most common measure of
dispersion is the classical variance, whose value decreases as the
number of data points included in the calculations, increases. Unfortunately
this is only true in the case of truly random processes, where the
variance of the mean decreases with the number N of data points as
var(X)=$\frac{\sigma^{2}}{N}$.

Allan stated that in the case of autocorrelated data the above equation
does not apply since there exists a possibility that the estimated
variance will diverge as the number of data points increases {[}12{]}.
So, we create a high-pass filter by extracting each k+1 value from
its previous one k and thus we remove any possible trends, fast fluctuations
or other peculiar characteristics. The Allan variance \textgreek{sv}$_{y}$$^{2}$(\textgreek{t})
is estimated at time intervals \textgreek{t}=m\textgreek{t}$_{0}$,
where $\tau_{0}$ is a minimum sampling time and m is usually chosen
to denote powers of two. 

$ $
\begin{equation}
\sigma_{y}^{2}(\tau)=\frac{1}{2(N-1)}\sum_{k=1}^{N-1}(y_{k+1}(\tau)-y_{k}(\tau))^{2}
\end{equation}

From the resulting plot of \textgreek{sv}$_{y}$$^{2}$(\textgreek{t})-\textgreek{t}
one can estimate the cut-off value after which the inclusion of more
data points does not lead to the decrease of the variance. In a log-log
plot the Allan variance is proportional to \textgreek{t}$^{\mu}$$ $
and \textgreek{m} = -\textgreek{a}-1, where \textgreek{a} is the spectral
exponent appearing in equation (3). Special cases {[}6,7{]} are: \textgreek{m}=1
(random walk noise), \textgreek{m}=0 (flicker noise) and \textgreek{m}=-1
(white noise).

\subsection{Modified Allan Variance}

This variance was first introduced in the domain of optics because
it divides white phase noise from flicker phase noise by the different
dependence on gate time. More specifically, ModVar is proportional
to \textgreek{t}$^{\lyxmathsym{\textminus}3}$ for the former and
\textgreek{t}$^{\lyxmathsym{\textminus}2}$ for the latter and ModStd
is proportional to \textgreek{t}$^{\lyxmathsym{\textminus}3/2}$ and
\textgreek{t}$^{\lyxmathsym{\textminus}1}$ correspondingly {[}13{]}.
Given an infinite sequence \{x$_{k}$$ $\} of samples evenly spaced
in time with sampling period \textgreek{t}$_{0}$$ $, the ModVar
is defined as

\begin{equation}
Mod\sigma_{y}^{2}(\tau)=\frac{1}{2n^{4}\tau_{0}^{2}(N-3n+1)}\sum_{j=1}^{N-3n+1}[\sum_{i=j}^{n+j-1}(x_{i+2n}(\tau)-2x_{i+n}(\tau)+x_{i}(\tau))]^{2}
\end{equation}

where n=1,2...N/3. The MVar obeys to a power law of the observation
time \textgreek{t} of the form $Mod\sigma_{y}^{2}\sim k\tau^{\mu}$
where $\mu=-3-a$ {[}14{]}.

\section{Results}

After removing the deterministic component (linear trend because of
the ambient temperature) of the data with the application of a first
difference filter, the resulting series is carefully examined for
correlations and noise. The lag plot and autocorrelation function
of the first-differenced time series are presented below, in fig.
(1) and fig. (2) correspondingly.

\noindent %
\begin{minipage}[t]{0.45\columnwidth}%
\noindent \begin{center}
\includegraphics[clip,width=7cm,height=5cm]{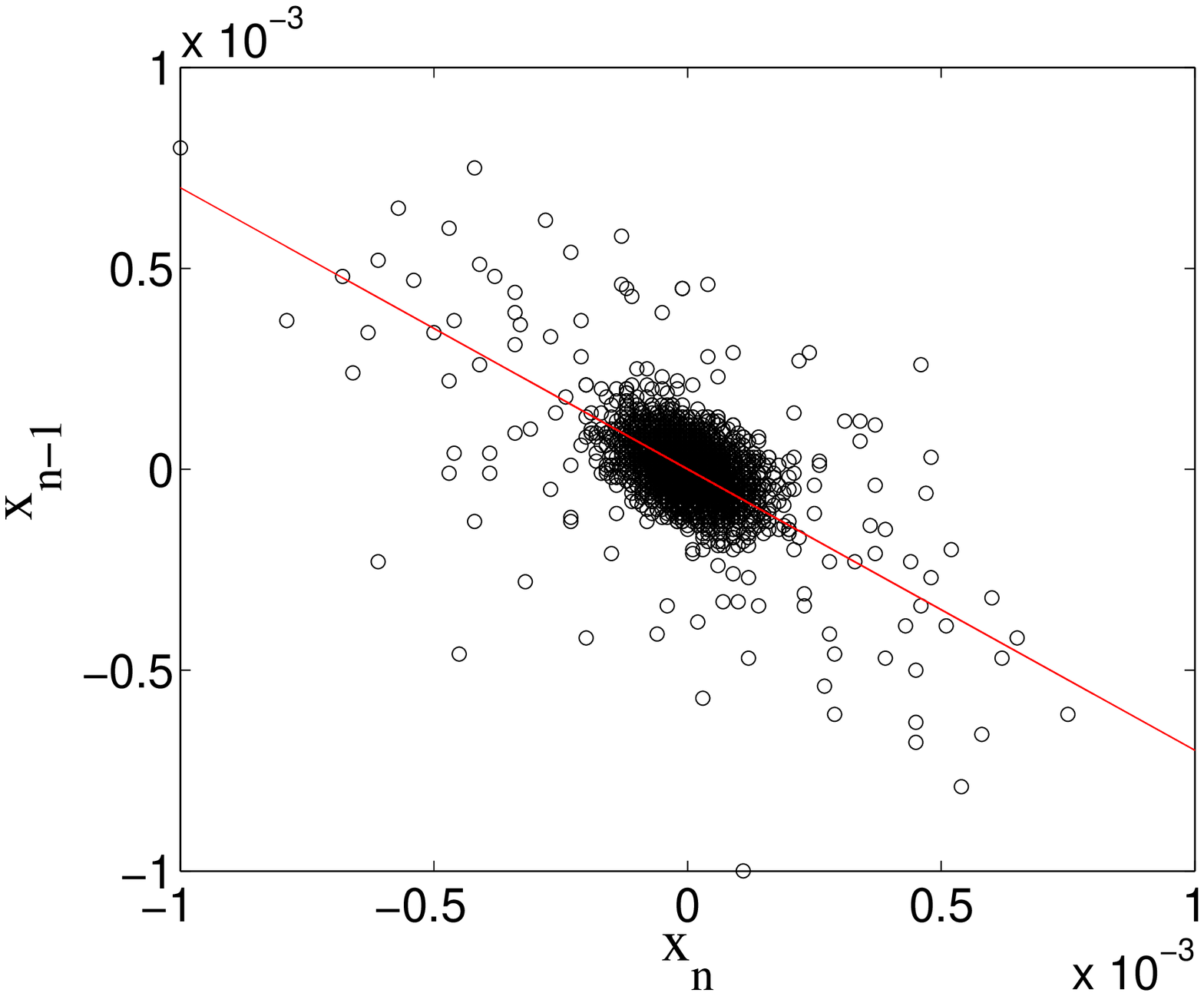}
\par\end{center}

\noindent \begin{center}
Figure 1. Lag Plot with negative correlations.
\par\end{center}%
\end{minipage}\hfill{}%
\begin{minipage}[t]{0.45\columnwidth}%
\noindent \begin{center}
\includegraphics[clip,width=7cm,height=5cm]{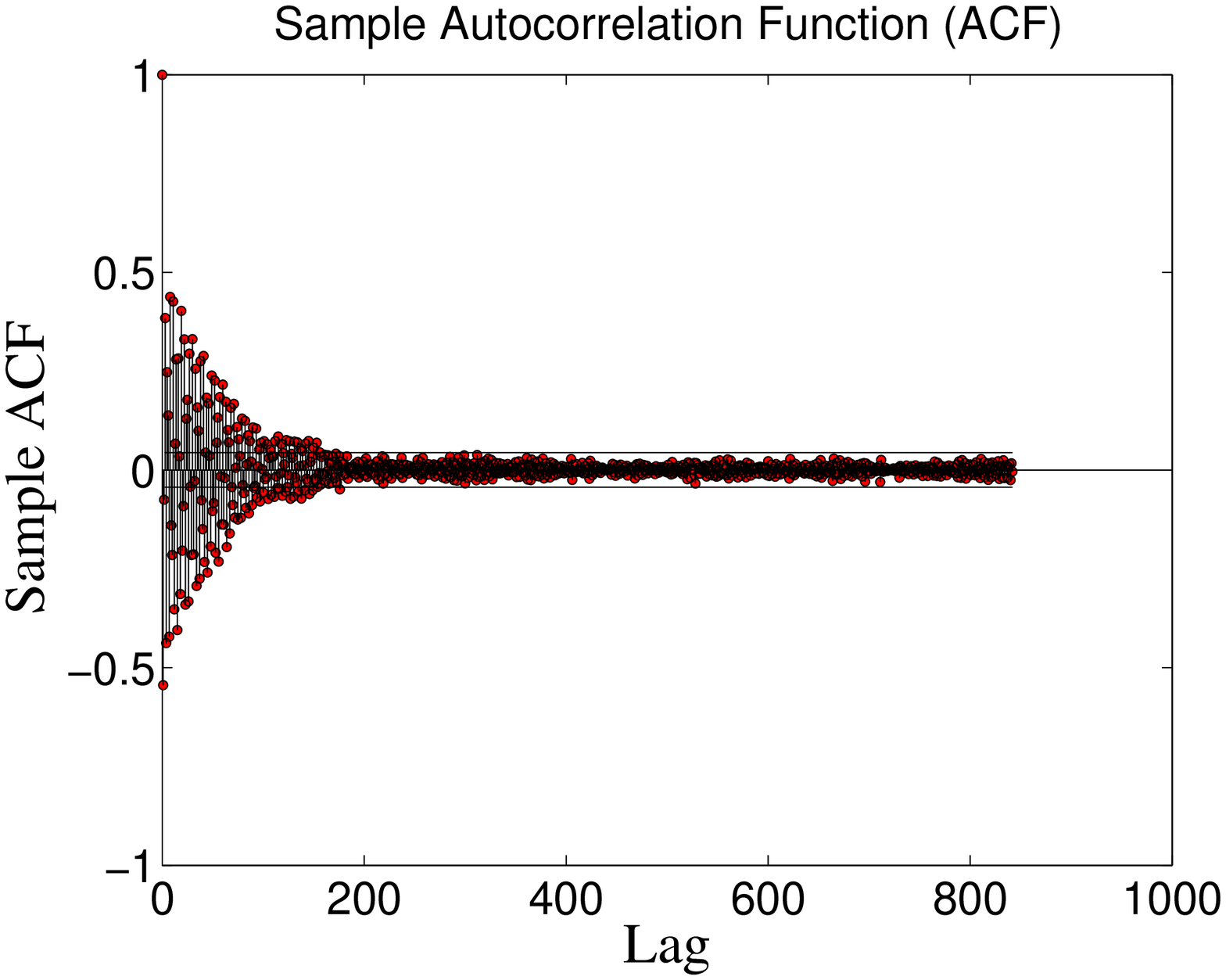}
\par\end{center}

\noindent \begin{center}
Figure 2. Autocorrelation function with oscillatory behavior.
\par\end{center}%
\end{minipage}

The lag plot reveals negative correlations between successive data
points with lag=1 but is unable to tell between White and Flicker
PM Noise. Moreover, the autocorrelation function changes between positive
and negative values for the first 200 lags (strong correlations) and
afterwards, the data become statistically independent falling well
within the 95\% confidence bands of $\pm\frac{2}{\sqrt{N}}$, implying
that the white noise model is now sufficient to describe the measurements.
In order to determine safely whether or not to use Allan Variance,
additional information is needed from the frequency domain analysis.
The Schuster periodogram and the Power Spectrum of the data are shown
in the following figures.

\noindent %
\begin{minipage}[t]{0.45\columnwidth}%
\noindent \begin{center}
\includegraphics[clip,width=7cm,height=5cm]{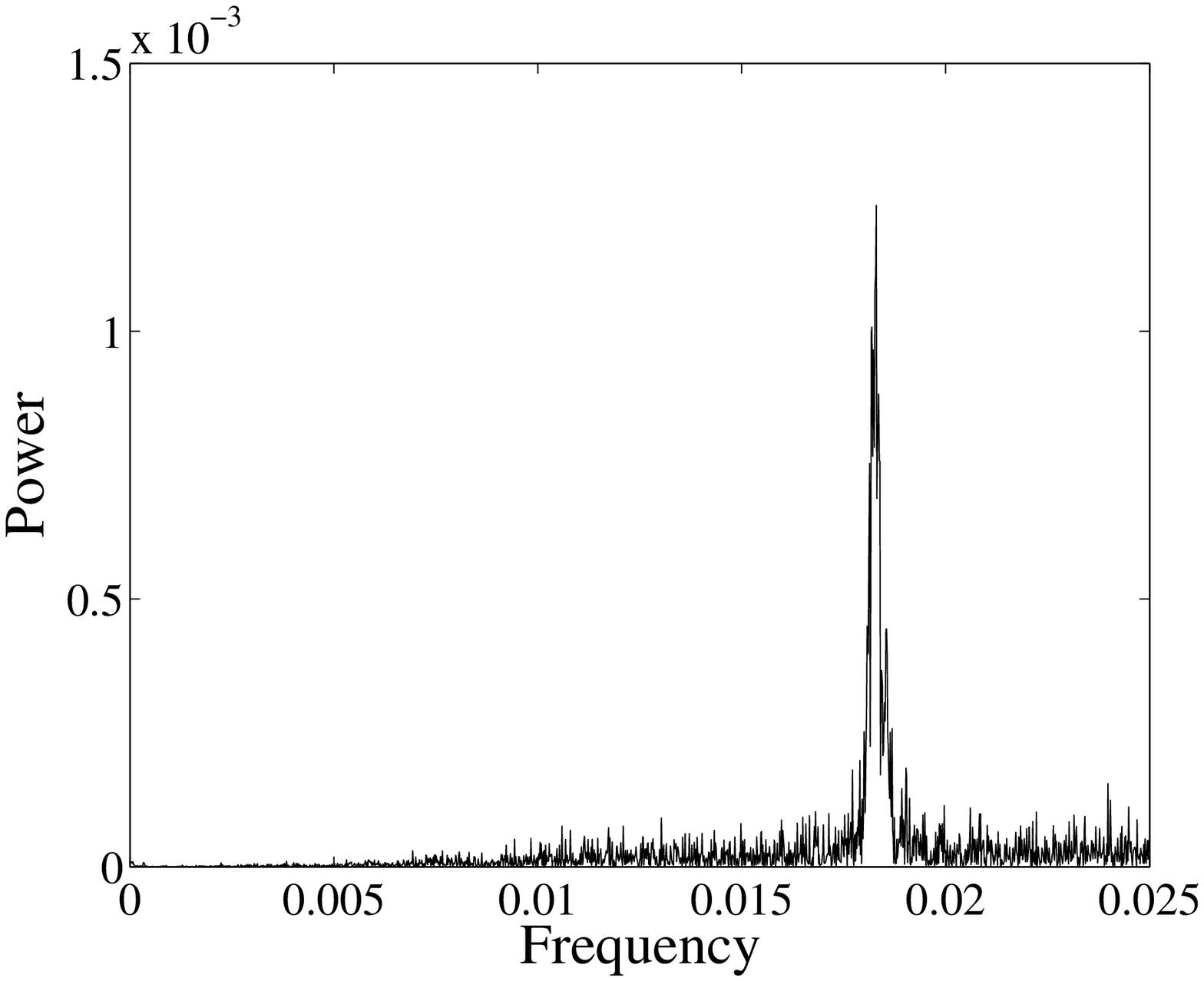}
\par\end{center}

\noindent \begin{center}
Figure 3. Periodogram with a single peak.
\par\end{center}%
\end{minipage}\hfill{}%
\begin{minipage}[t]{0.45\columnwidth}%
\noindent \begin{center}
\includegraphics[clip,width=7cm,height=5cm]{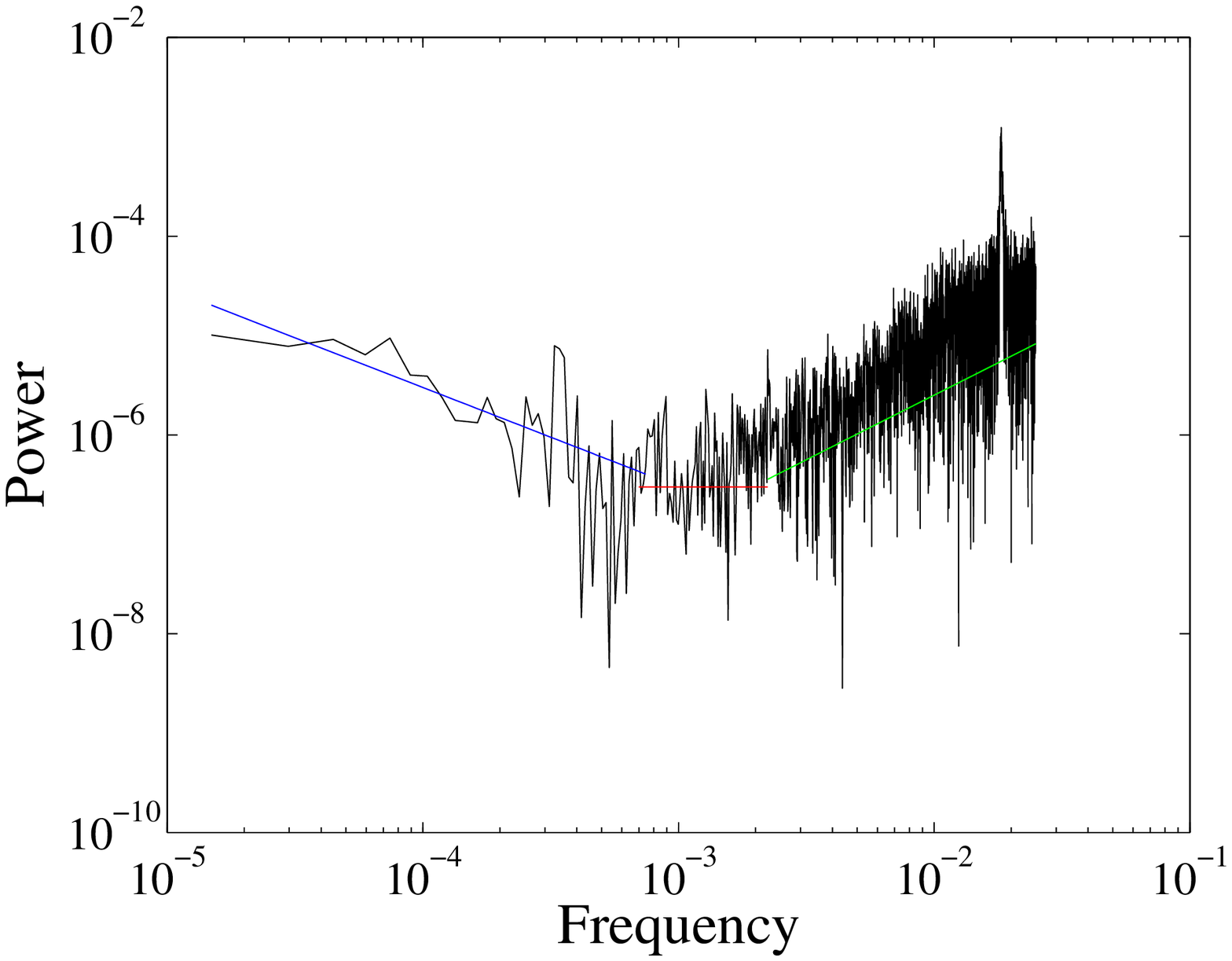}
\par\end{center}

\noindent \begin{center}
Figure 4. Power Spectrum with three different noise regions.
\par\end{center}%
\end{minipage}

$\vphantom{}$$\vphantom{}$$\phantom{}$

The periodogram reveals a strong oscillation at approximately 0.0183
Hz (or 54.64 sec). The power spectrum in a log-log scale, seems to
be divided in three regions: at lower frequencies the blue line indicates
a power law model of the form 1/f (Flicker FM Noise), the red line
indicates that the White FM Noise model is adequate to describe the
data and finally at higher frequencies the green line indicates a
power law model 1/f$^{-1}$$ $ or 1/f$^{-2}$ (PM Noise). Thus, there
exist two cut-off frequencies at which evident changes in slope occur,
revealing the different noise areas. The total power spectrum is given
by the equation Sy(f) = h$_{0}$f$_{0}$ + h$_{-1}$f$_{-1}$ + h$_{1}$f$_{1}$
(or h$_{2}$f$_{2}$), where h$_{i}$ are the intensity coefficients.
In order to be sure for the oscillation that FFT revealed, we apply
the Bretthorst methodology and according to theory we expect a single
spike at the frequency of 0.0183 Hz. In order to keep the same quantities
of measurement we will calculate P(f|D,I). As can be observed in fig.
(5), the single spike (red line) confirms the existence of f$_{osc}$=
0.0183 Hz and as predicted by theory all the other information of
the periodogram (black diagram) is suppressed to a uniform plateau
of zero slope.

\noindent \begin{center}
\includegraphics[clip,width=7cm,height=5cm]{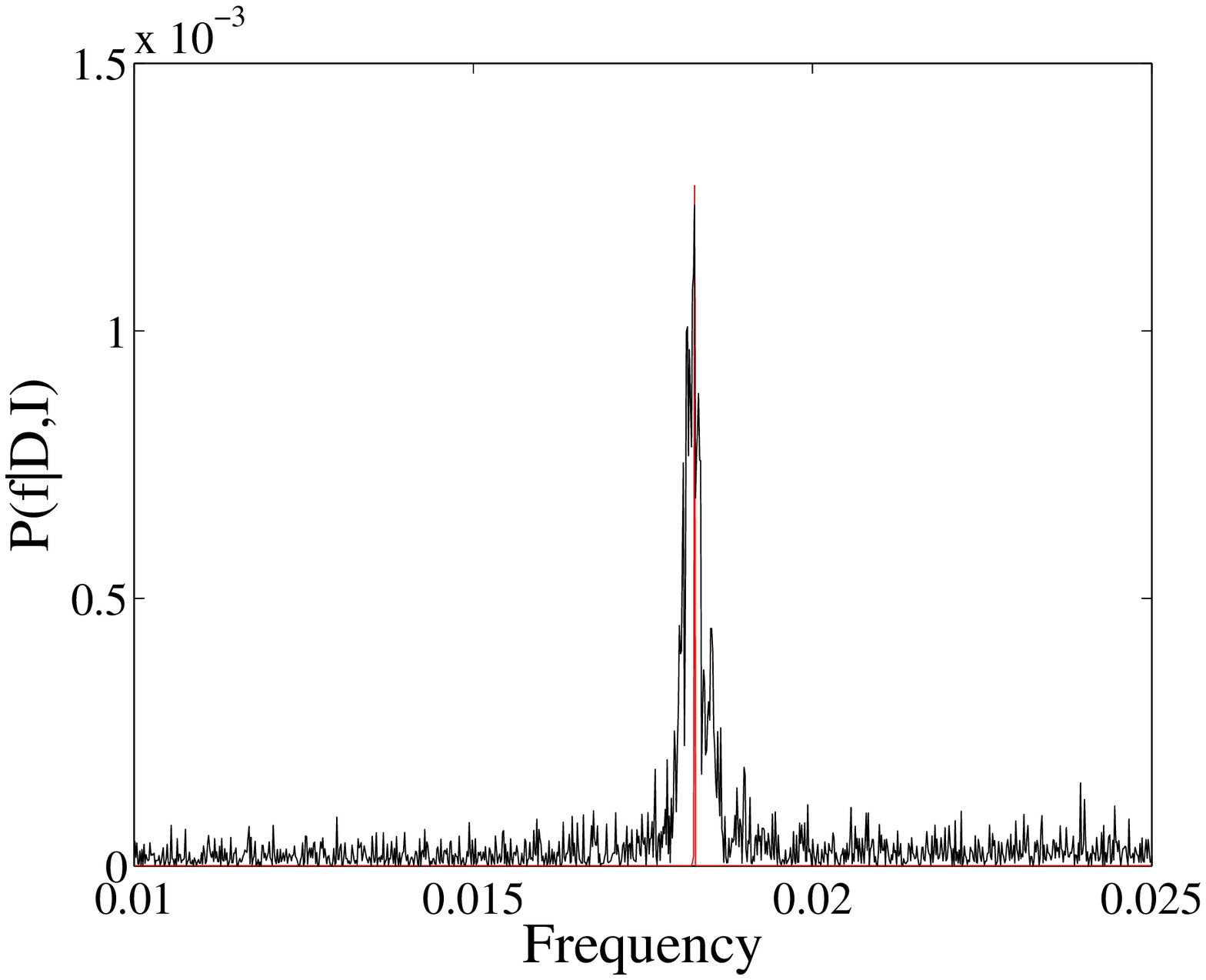}
\par\end{center}

\noindent \begin{center}
Figure 5. Comparison of Schuster and Bretthorst periodogram
\par\end{center}

From all the above, it is obvious that the use of classical variance
for the calculation of type A uncertainty is inadequate and Allan
variance will be used instead. In addition, Modified Allan variance
will give a straightforward answer for the phase noise characterization
problem in our data set. Fig. (6) below, gives the dependence of the
Allan deviation over time. Clearly, there exist three different regions,
in accordance with the power spectrum of fig. (4): the phase noise
region \textgreek{m}=-1 (or \textgreek{m}=-2 for the Allan Variance)
indicated by the blue line, the white FM noise area \textgreek{m}=-0.5
(or \textgreek{m}=-1 for the Allan variance) adjusted by the red line
and finally the flicker FM floor (\textgreek{m}=0). For \textgreek{t}=256,
\textgreek{sv}(\textgreek{t})=1.9{*}10$^{-6}$ in contrast to classical
std which estimates \textgreek{sv}=9.57{*}10$^{-5}$. Note that the
last two points are not included in the calculation. Probable causes
of the flicker floor are power supply voltage fluctuations, changes
in the components of the artifact and microwave power changes. Fig.
(7) gives the dependence of the Modified Allan deviation over time.
The slope of the blue fitted line is -3/2, which indicates that white
PM Noise appears at small times and correspondingly high frequencies,
as already observed in the power spectrum. 

\noindent %
\begin{minipage}[t]{0.45\columnwidth}%
\noindent \begin{center}
\includegraphics[clip,width=7cm,height=5cm]{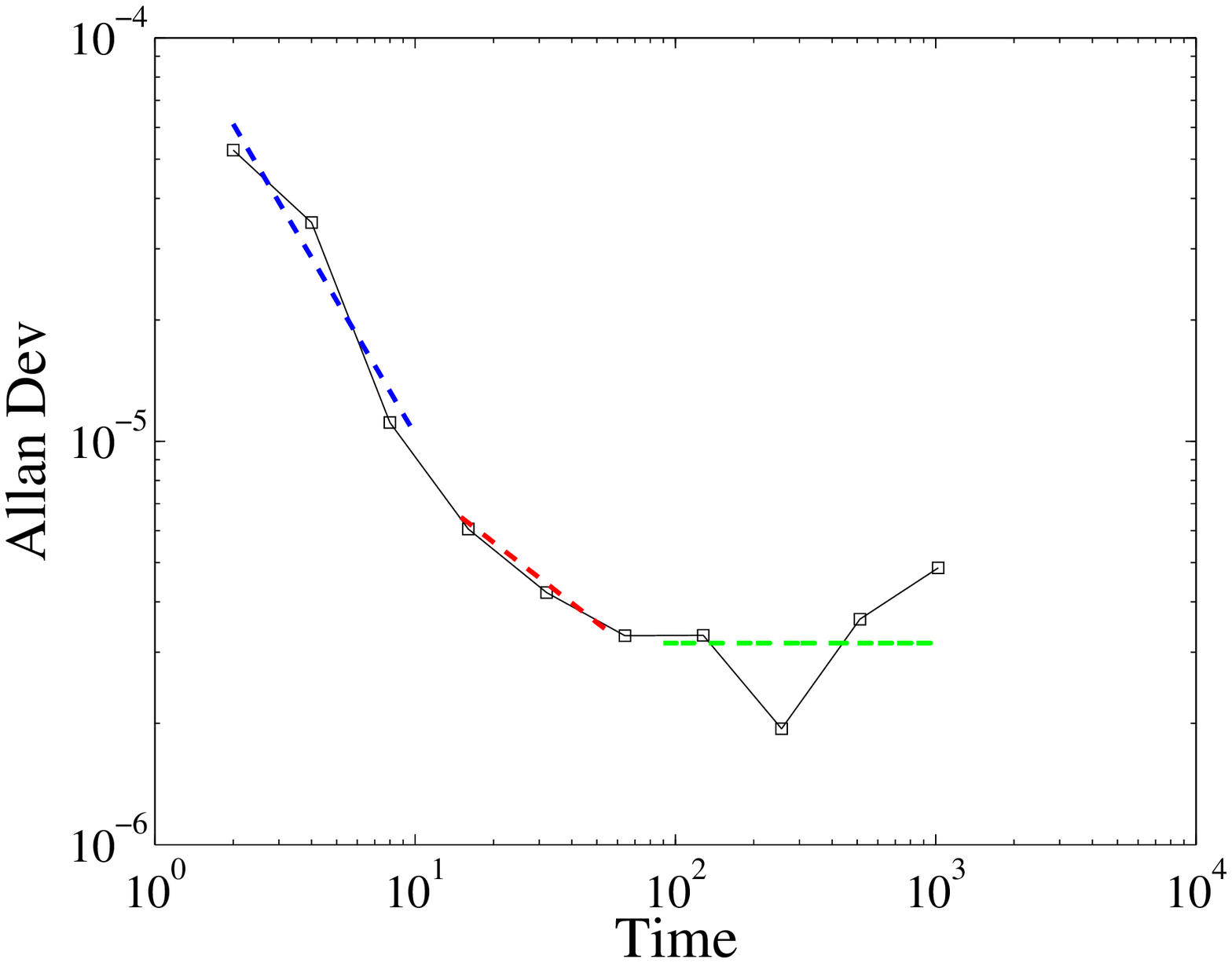}
\par\end{center}

\noindent \begin{center}
Figure 6. Allan variance.
\par\end{center}%
\end{minipage}\hfill{}%
\begin{minipage}[t]{0.45\columnwidth}%
\noindent \begin{center}
\includegraphics[clip,width=7cm,height=5cm]{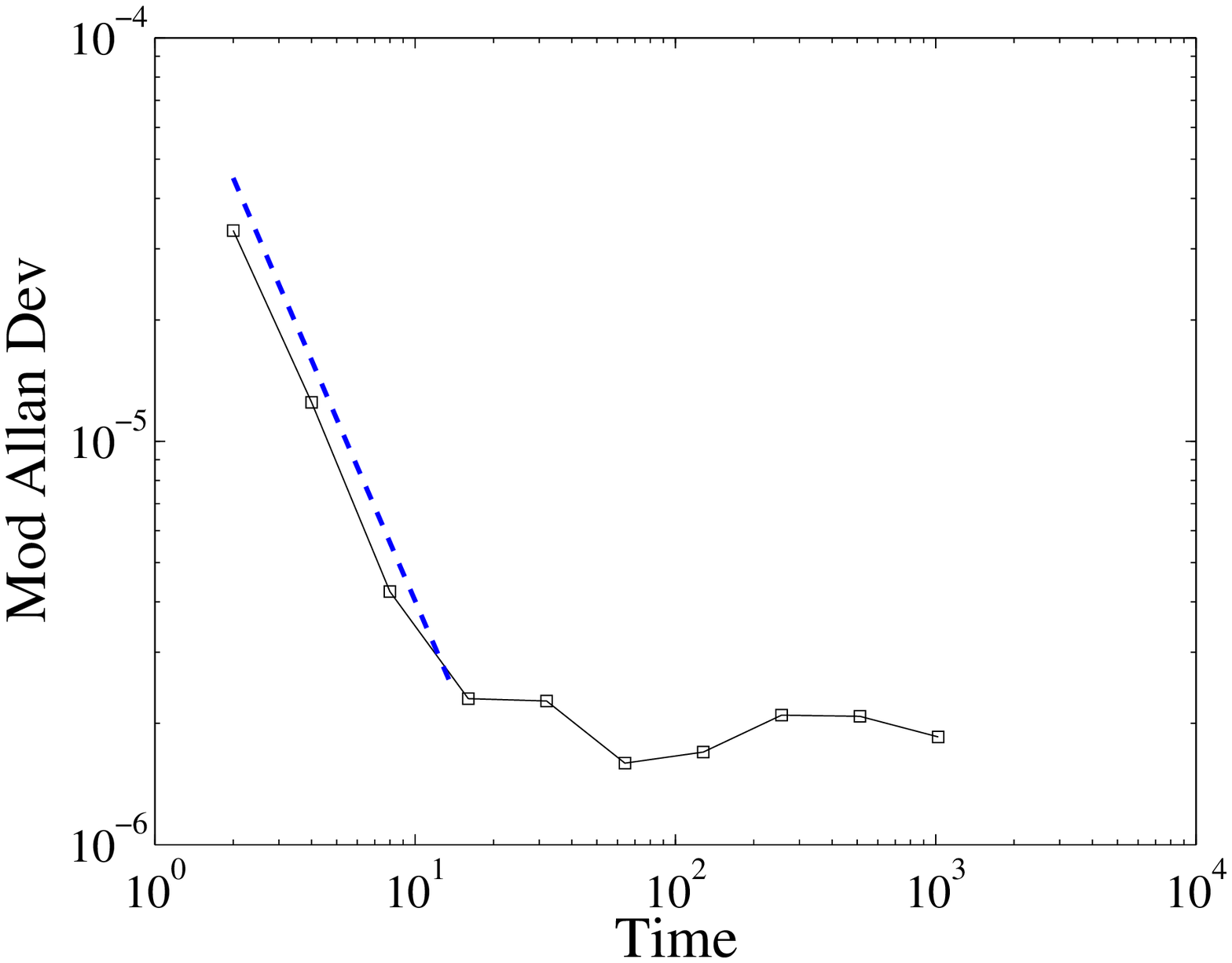}
\par\end{center}

\noindent \begin{center}
Figure 7. Modified Allan variance.
\par\end{center}%
\end{minipage}

\section{Conclusions}

The main purpose of this work is to point out the importance of considering
measurement results as time series and of using the time series analysis
methods to test for correlations and underlying stochastic noise.
Only in the absence of correlation the use of the expression $\sigma/\sqrt{N}$
is appropriate to characterize the random uncertainty. Otherwise,
the Allan variance is an alternate, more appropriate tool to characterize
the dispersion of a set of experimental results. 

In the case of phase noise, only Modified Allan variance provides
accurate results and discriminates flicker from white PM noise. Bretthorst
periodogram is a powerful tool of bayesian spectral analysis which
detects oscillations in an output signal and identifies peaks at a
very close distance. For the case study of mass measurements we consider
here, all the above techniques give consistent results.


\begin{thebibliography}{10}
\bibitem{key-1}N.F. Zhang, Calculation of the uncertainty of the
mean of autocorrelated measurements, \textit{Metrologia,} 43, S276-S281
(2006).

\bibitem{key-2}N.J. Kasdin, Discrete simulation of colored noise
and stochastic processes and 1/f$^{a}$ power law noise generation,
\textit{Proceedings of the IEEE,} 83 (5), 802-827 (1995).

\bibitem{key-3}A. Hajimiri and T.H. Lee, A general theory of phase
noise in electrical oscillators, \textit{IEEE Journal of Solid-State
Circuits}, 33 (2), 179-194 (1998).

\bibitem{key-4}E. Rubiola and V. Giordano, Oscillators and the characterization
of frequency stability: an introduction, Phase noise metrology in
\textit{Noise, Oscillators and Algebraic Randomness (M. Planat, eds.),
Springer, Chapelle des Bois France, 1999, 175-215.}

\bibitem{key-5}D.A. Howe et al, Vibration-Induced PM Noise in Oscillators
and Measurements of Correlation with Vibration Sensors in \textit{Proceedings
of the 36th Annual Precise Time and Time Interval (PTTI) Systems and
Applications Meeting}, (L. Breakiron ed.), Washington D.C., 2005,
494-498

\bibitem{key-6}D.W. Allan, Should the Classical Variance Be Used
As a Basic Measure in Standards Metrology?, \textit{IEEE Transactions
on Instrumentation and Measurement}, IM-36, No 2, 646-654 (1987).

\bibitem{key-7}T.J. Witt, Testing for Correlations in Measurements,
\textit{Advanced Mathematical and Computational Tools in Metrology
IV}, World Scientific Publishing Company, 273-288 (2000).

\bibitem{key-8}D.W. Allan and J.A. Barnes, A Modified \textquotedbl{}Allan
Variance\textquotedbl{} with Increased Oscillator Characterization
Ability, \textit{Proceedings of the 35th Annual Frequency Control
Symposium}, 470-475 (1981).

\bibitem{key-9}G.L. Bretthorst, An introduction to parameter estimation
using Bayesian Probability Theory in \textit{Maximum entropy and Bayesian
methods}, (P.F. Fougere ed.), Kluwer Academic Publishers, The Netherlands,
1990, 53-79.

\bibitem{key-10} G.L. Bretthorst, \textit{Bayesian Spectrum Analysis
and Parameter Estimation}, (J. Kimmel ed.), Springer-Verlag, Berlin
Heidelberg, 1988.

\bibitem{key-11}A. Gelman et al, \textit{Bayesian Data Analysis},
(2nd ed), Chapman \& Hall/CRC, Boca Raton, Florida, 2004.

\bibitem{key-12}J. Jespersen, Introduction to the time domain characterization
of frequency standards, \textit{Tutorials from the Annual Precise
Time and Time Interval (PTTI) applications and Planning Meeting} (23rd),
Pasadena, California, 1991.

\bibitem{key-13}Won-Kyu Lee et al, The uncertainty associated with
the weighted mean frequency of a phase-stabilized signal with white
phase noise, \textit{Metrologia}, 47, 24-32 (2010).

\bibitem{key-14}S. Bregni, The Modified Allan Variance as Time-Domain
Analysis Tool for Estimating the Hurst Parameter of Long-Range Dependent
Traffic, \textit{Proceedings of the IEEE Global Telecommunications
Conference}, 3, 1406-1410 (2004).\end{thebibliography}
\end{document}